\documentclass[prl,aps,twocolumn,showpacs,home,floats]{revtex4}
\usepackage{graphics}
\begin{document}

\title{{\rm\small\hfill (submitted to Phys. Rev. Lett.)}\\  
First-principles, atomistic thermodynamics for oxidation catalysis}

\author{Karsten Reuter}
\author{Matthias Scheffler}
\affiliation{Fritz-Haber-Institut der Max-Planck-Gesellschaft, Faradayweg 4-6, D-14195 Berlin, Germany}
\affiliation{{\rm (Received 16 October)}}

\begin{abstract}
Present knowledge of the function of materials is largely based on
studies (experimental and theoretical) that are performed at low
temperatures and ultra-low pressures. However, the majority of
everyday applications, like e.g. catalysis, operate at atmospheric
pressures and temperatures at or higher than 300~K. Here we employ
{\em ab initio}, atomistic thermodynamics to construct a phase diagram
of surface structures in the $(T,p)$-space from ultra-high vacuum to
technically-relevant pressures and temperatures. We emphasize the value
of such phase diagrams as well as the importance of the reaction {\em kinetics}
that may be crucial e.g. close to phase boundaries.
\end{abstract}

%\date{Received 16 October 2002}     


% 82.65.Mq  Heterogeneous catalysis at surfaces, surf. chemistry
% 68.35.Md  Surface thermodynamics, surface energies
% 68.43.Bc  Ab initio calculation of adsorbate structure and reactions

\maketitle
The main prerequisites for reaching a microscopic understanding of
heterogeneous catalysis are the identification of the composition
and geometry of the catalyst' surface and the determination of
the various chemical reactions that take place under realistic
conditions. Unfortunately, most {\em Surface Science} experimental
techniques are difficult if not impossible to use at the pressures (of
the order of one atmosphere) and temperatures (often higher than 300~K)
that are typically applied in steady-state catalysis. Therefore, what is
considered to be important elementary processes at the catalyst' surface
(e.g. dissociation, diffusion, and chemical reactions) has usually been
concluded from chemical intuition and extensive knowledge from ultra-high
vacuum (UHV) experiments. However, several studies revealed the danger of
this approach (see e.g. Ref. \cite{Stampfl02} and references therein),
and the need to reliably bridge the temperature and pressure gap between 
UHV and ``real life'' is probably the main challenge in modern
{\em Surface Science}.

In {\em ab initio} theory the consideration of high temperature and
high pressure can be achieved by explicitly taking into account the surrounding
gas phase in terms of ``{\em ab initio}, atomistic thermodynamics'' (cf.
e.g. Refs. \cite{Weinert86,Scheffler87,Kaxiras87,Qian88,Reuter02}).
This is also an appropriate (first) approach to steady-state catalysis,
which is often run close to thermodynamic equilibrium (or a constrained
equilibrium) to prevent catalyst degradation. We will qualify this
statement below when analyzing our results. In the following we show how
the combination of thermodynamics and density-functional theory (DFT)
can be applied to obtain the lowest-energy surface structures in
a (constrained) equilibrium with the surrounding gas phase,
thus enabling us to construct a $(T,p)$-diagram of the stability
regions (or metastability regions) of different surface phases.  We
use the example of CO oxidation over a RuO$_2$(110) model catalyst to 
illustrate the concepts and conclusions that can be derived from such
an {\em ab initio} surface phase diagram.

At given temperature, $T$, and partial pressures, $\{p_i\}$, the stable
surface structure is the one that has the lowest surface energy, 
$\gamma(T,\{p_i\}) = 1/A  [ G(T,\{p_i\}) - \sum_i N_i \mu_i(T, p_i) ]$. 
Here, $G(T,\{p_i\})$ is the Gibbs free energy of the finite crystal, $N_i$
and $\mu_i(T, p_i)$ are number and chemical potential of the species of the
$i$th type, and $\gamma(T,\{p_i\})$ is measured in energy
per unit area by dividing through the surface area, $A$. The surface free
energy is thus a function of the chemical potentials of all species in the 
system, e.g. in the present application to CO oxidation over RuO$_2$(110)
we have three $\mu_i(T, p_i)$ corresponding to the three chemical elements: 
Ru, C, and O. Considering that the surface is not only in equilibrium with
the gas phase, but also in equilibrium with the underlying metal oxide, 
implies that the chemical potentials of Ru and O are not independent 
variables but tied by the Gibbs free energy of the bulk oxide. As a 
consequence, the surface free energy depends only on two chemical 
potentials, which we choose to be those of O and CO. Because chemical
potentials are directly related to temperature and partial pressure,
a comparison to the whole range of experimentally accessible gas phase
conditions is achieved.\cite{Reuter02,comment-1} Here, we note that
the above description refers to a {\em constrained} thermodynamic 
equilibrium only with the reactants, because it assumes that the 
O$_2$ and CO gas phase are independent {\em reservoirs}, and the direct
(non catalytic) formation of CO$_2$ in the gas phase can be ignored 
because of its negligible rate.

As shown previously \cite{Reuter02}, for RuO$_2$ it is a good approximation
to assume that the vibrational contributions to the Gibbs free energy, 
$G(T,\{p_i\})$, nearly equal the corresponding terms of the bulk oxide.
We will therefore replace these quantities by the corresponding total
energies, which are calculated using DFT. We use the generalized gradient 
approximation (GGA) for the exchange-correlation functional \cite{Perdew96},
and the all-electron Kohn-Sham equations are solved by employing the 
full-potential linear augmented plane wave (FP-LAPW) method 
\cite{Wien,parameters}. The high accuracy of these calculations (cf. 
Ref. \cite{Reuter02}) implies that the numerical accuracy of the relative
$\gamma(T,\{p_i\})$ is better than $\pm 5 $~meV/\AA$^2$. We also note that
using the local-density approximation (LDA) as exchange-correlation 
functional gives essentially the same result for the surface phase diagram
(cf. Fig. \ref{figure-phase-diagram}). Thus, the DFT accuracy for this 
system's surface phase diagram is very high, partly due to the fact that
it follows only from energy {\em differences}.

\begin{figure}[t!]
\scalebox{0.65}{\includegraphics{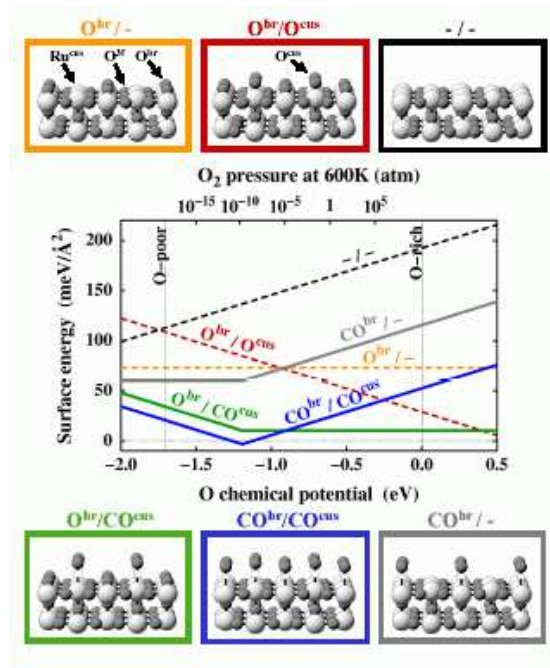}}
\caption{\label{figure-gamma(mu-O)}
Surface free energies, $\gamma (T,\{p_i\})$, in the experimentally 
accessible range of the oxygen chemical potential.\cite{Reuter02,comment-1}
Shown are three RuO$_2$(110) terminations (dashed lines and black, red,
and orange insets), and the three most stable CO containing surface 
phases for the CO-rich limit (solid lines and gray, green, and blue 
insets). The labels note whether bridge or cus sites are occupied by O
or CO, or are empty $(-)$. Ru atoms are drawn as light, large, O atoms 
as dark, small, and C atoms as white, small spheres.}
\end{figure}

\begin{figure}[t!]
\scalebox{0.65}{\includegraphics{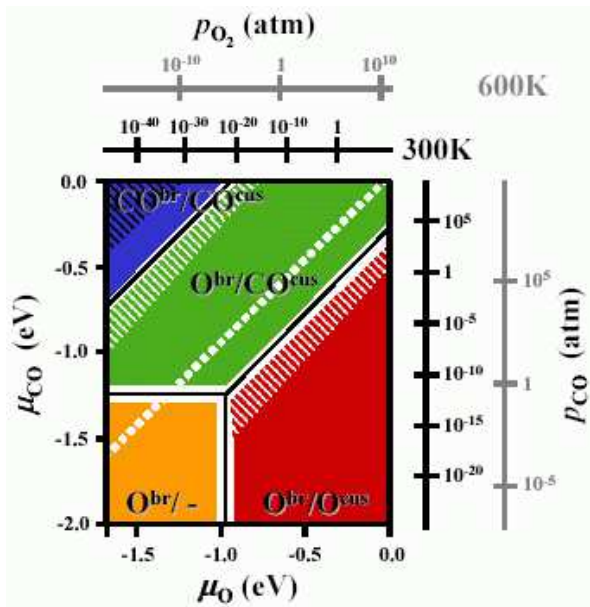}}
\caption{\label{figure-phase-diagram}
Regions of the lowest-energy structures in ($\mu_{\rm O}$, 
$\mu_{\rm CO}$)-space (see Fig. \ref{figure-gamma(mu-O)} for
the nomenclature). The additional axes give the corresponding
pressure scales at $T= 300$~K and $600$~K. In the blue-hatched 
region gas phase CO is transformed into graphite. Regions that
are particularly strongly affected by kinetics are marked by 
white hatching (see text).}
\end{figure}

Ruthenium is well known to exhibit an exceptionally high activity for the
catalytic oxidation of CO, with turnover rates exceeding those of the 
more commonly employed materials like Rh, Pd, or Pt.\cite{Cant78,Peden86}
For Ru(0001) this high activity could be related to RuO$_2$(110) oxide
patches that are formed in the reactive environment.\cite{Boettcher97,Over00,Kim01}
The atomic arrangement at RuO$_2$(110) can be described by starting
from the stoichiometric surface (see the orange-framed inset in Fig. 
\ref{figure-gamma(mu-O)}, labeled O$^{\rm br}/-$). There are two 
different oxygen atoms at this surface: threefold-bonded O${}^{\rm 3f}$
(as is an O atom in bulk RuO$_2$) and twofold bridge-bonded O${}^{\rm br}$.
Likewise, there are two different surface Ru atoms, one of which
is bonded to six oxygen neighbors (as is a Ru atom in bulk RuO$_2$),
and one is bonded to only five O neighbors. The latter represents a
``coordinatively unsaturated site'' and is called Ru$^{\rm cus}$.
The bulk stacking sequence would be continued by adsorbing additional oxygen
atoms on top of the Ru$^{\rm cus}$ atoms. This gives the red-framed
oxygen-rich geometry, O$^{\rm br}$/O$^{\rm cus}$. The adsorption energy
for a single O atom on the Ru$^{\rm cus}$ atom of the O$^{\rm br}/-$ surface
is calculated to be exothermic by 1.2 eV/adatom (at $\mu_{\rm O}= 0 $
eV)\cite{comment-3}. And the energy of an oxygen vacancy in the
O$^{\rm br}/-$ surface is calculated to be endothermic by 2.5~eV/vacancy 
(removing a O${}^{\rm br}$ atom) and $\approx 3.5$~eV/vacancy (removing
a O${}^{\rm 3f}$ atom) at $\mu_{\rm O}=0 $ eV\cite{comment-3}.
Removing all bridge-bonded oxygen atoms from the O$^{\rm br}/-$ surface 
gives the black-framed, oxygen-poor geometry $-/-$ in Fig. 
\ref{figure-gamma(mu-O)}, which represents the third possible
$(1 \times 1)$ termination of RuO${}_2$(110).

The $\mu_{\rm O}$ scale can be converted into a pressure scale at
any chosen temperature,\cite{Reuter02,comment-1} and this is noted
at the top of Fig. \ref{figure-gamma(mu-O)} for $T = 600$~K, a
typical annealing temperature used experimentally in this system 
\cite{Over00,Kim01}. From the computed surface free energies of the 
aforementioned three RuO$_2$(110) terminations, we obtain that even at
the lowest possible O$_2$ pressures, O atoms will at least occupy the 
bridge sites, leading to the stoichiometric surface termination 
(the orange line tagged as O$^{\rm br}/-$ in Fig. \ref{figure-gamma(mu-O)}). 
At even lower $\mu_{\rm O}(T, p_{\rm O_2} )$ than the O-poor boundary 
in Fig. \ref{figure-gamma(mu-O)}, RuO$_2$ will decompose into Ru
metal.\cite{Reuter02}

We now consider the presence of CO as a second species, discussing the
surface structure of RuO$_2$ in the constrained equilibrium with a CO
and O$_2$ gas phase. For this we computed what we believe are all
possible $(1 \times 1)$ surface phases including O and CO at the
surface.\cite{comment-4} Figure \ref{figure-gamma(mu-O)} shows the
corresponding surface energies in the CO-rich limit for $\mu_{\rm CO}$ 
of three CO containing structures that will play a role in the later 
discussion. In this CO-rich limit and if the O chemical potential is 
below $-1.2$~eV, CO is transformed into graphite and 1/2 O$_2$. This
is why the CO related (gray, blue, and green) lines in Fig. 
\ref{figure-gamma(mu-O)} exhibit a kink at $\mu_{\rm O} = -1.2$~eV. 
As apparent from Fig. \ref{figure-gamma(mu-O)} a phase with CO 
occupying all bridge sites (the gray line tagged as CO$^{\rm br}/-$)
is barely more stable than the hitherto discussed pure O-terminations;
and occupation of also the cus sites (CO$^{\rm br}$/CO$^{\rm cus}$) leads
to a (meta)stable phase only at low O chemical potential. Towards higher
$\mu_{\rm O}$ we find a third relevant geometry, that exhibits O atoms
at the bridge sites and CO at the cus sites, (O$^{\rm br}$/CO$^{\rm cus}$).

If we now in addition to $\mu_{\rm O}$ investigate the dependence on
$\mu_{\rm CO}$, Fig. \ref{figure-gamma(mu-O)} turns into a complicated
three-dimensional plot. We therefore present our results by showing only 
the lowest-energy surface structures (concluded from many calculations of 
the type shown in Fig. \ref{figure-gamma(mu-O)} for any value of 
$\mu_{\rm CO}$). Figure \ref{figure-phase-diagram} displays this central
result of our study from which the stability regions of all surface 
phases in (constrained) equilibrium with an environment formed by O$_2$ 
and CO can be seen. At very low CO chemical potential corresponding to 
very low CO pressure in the gas phase, we recover our results obtained
for RuO$_2$(110) just in equilibrium with an O$_2$ environment (dashed 
lines in Fig. \ref{figure-gamma(mu-O)}): At low O chemical potential we
find the stoichiometric termination with O at bridge sites, while towards 
higher O$_2$ pressures the O-rich termination with O additionally located 
at the cus sites becomes more stable. Increasing the CO content in the
gas phase, CO is first bound at the cus sites. This is easier at low O 
chemical potentials, where these sites are free. At higher $\mu_{\rm O}$ 
oxygen atoms also compete for cus sites, which is why the 
O$^{\rm br}$/CO$^{\rm cus}$ phase becomes only more stable than the 
O$^{\rm br}$/O$^{\rm cus}$ phase at progressively higher CO chemical 
potentials, cf. Fig. \ref{figure-phase-diagram}. Finally, at a high CO 
and low O chemical potential in the upper-left corner of Fig. 
\ref{figure-phase-diagram}, we find that CO is also stabilized at the 
bridge sites (instead of O), leading to a completely CO covered surface
(CO$^{\rm br}$/CO$^{\rm cus}$).

The knowledge of such a phase diagram (Fig. \ref{figure-phase-diagram}) 
is a first, important step towards an understanding where in 
$(T,\{p_i\})$-space catalysis may be most efficient, but in parallel it
also allows to systematically discuss under which gas phase conditions
our constrained thermodynamic equilibrium approach may break down:
We recall that the imposed constraints are that $(i)$ in the gas phase the two 
molecules will not react to CO$_2$, which is appropriate because the
free-energy barrier is significant. Thus, CO$_2$ is only formed at the 
catalyst surface from where it will desorb immediately. Furthermore,
it is assumed that $(ii)$ the CO$_2$ formation at the surface is slower
than other adsorption and desorption processes, so that the surface
can maintain its equilibrium with the reactant gas phase. Finally,
$(iii)$ the gas phase should not affect the stability of the RuO${}_2$
bulk structure. 

With respect to the latter point, we note that for $\mu_{\rm CO} > \mu_{\rm O} 
+ \Delta$, RuO${}_2$ will decompose into Ru metal ($\Delta^{\rm th} = 
0.06$ eV, see the dotted white line in Fig. \ref{figure-phase-diagram}; 
and $\Delta^{\rm exp} = 0.26$ eV\cite{JANAF}). Thus, a noticeable part
of Fig. \ref{figure-phase-diagram} (the blue and most of the upper-left 
green region) refers to metastable situations and will not prevail for 
long under realistic conditions. As for the remaining part of the
phase diagram, it is clear from the pressure scales in Fig. 
\ref{figure-phase-diagram} that the stoichiometric surface (orange region,
denoted as O$^{\rm br}/-$), that had been observed in UHV experiments, is 
not very important under high-pressure, catalytically relevant conditions. 
Obviously, for oxidation catalysis it is important that both reactants 
(here CO and O) are adsorbed at the surface. In this respect, the green 
region (O$^{\rm br}$/CO$^{\rm cus}$) appears to be the most relevant one
at first glance, and we have to ask if our remaining assumption $(ii)$
of the constrained thermodynamic approach is appropriate here: If the 
reaction CO$^{\rm cus}$ + O$^{\rm br}$ $\rightarrow$ CO$_2$ were faster 
than the dissociative adsorption of O$_2$, adsorbed CO would rapidly 
eat away bridge-bonded oxygen atoms, and surface kinetic effects would 
render the oxygen concentration much lower than the full monolayer that 
is behind the atomic structure of the green region. Fortunately, for 
this region the CO$_2$ formation energy barrier is calculated to be 
noticeable (1.2 eV, in agreement with Ref. \cite{Liu01}), and the energy 
to create vacancies in the O-bridge layer is high (see above). 
Correspondingly, the thermodynamic approach seems to be valid here.

The question remains where in $(T,p)$-space do we expect the RuO$_2$
catalysis to be most efficient? In this context we note that a so-called
stable phase is not stable on an atomistic scale, but represents an
average over many processes such as dissociation, adsorption, diffusion,
association and desorption. As all these processes and their interplay 
\cite{Stampfl02} are of crucial importance for catalysis, regions in
$(T,p)$-space where such fluctuations are particularly pronounced can be
expected to be most important. This is the case close to the boundaries
between different phases: At finite temperatures, the transition into a 
neighboring phase occurs not abruptly in ($\mu_{\rm O}$, $\mu_{\rm CO}$)-space,
but over a pressure range in which the other phase gradually becomes 
more populated. The resulting phase coexistence at the catalyst surface 
may then lead to a significantly enhanced dynamics, in which even 
additional reaction mechanisms can take place and/or (dynamic) domain pattern
formation may occur. Assuming a canonic distribution we estimated in Fig. 
\ref{figure-phase-diagram} the region on both sides of the boundaries in
which the respective other phase is present at least at a 10\% concentration
at room temperature (white areas). From the given pressure scales we see that of 
particular interest for catalysis is the green/red boundary between the 
O$^{\rm br}$/CO$^{\rm cus}$ and O$^{\rm br}$/O$^{\rm cus}$ phases, where 
oxygen atoms and CO molecules compete for the same site, i.e. the cus site. 

In fact, in this phase-transition region and in the neighboring hatched 
part of the red phase in Fig. \ref{figure-phase-diagram}, 
assumption $(ii)$ of our constrained equilibrium theory is not 
valid and the oxygen coverage will presumably be noticeably lower
than what the present approach suggests: Under these conditions
adsorbed CO will react with O$^{\rm cus}$, and the filling
of empty sites with CO will be fast, whereas the filling with oxygen atoms 
will be slow. In this region high catalytic activity is expected and we 
note that here the coverage and structure (i.e., the very dynamic behavior)
must be modeled by statistical mechanics. However, this is not the point of 
the present paper that aims to {\em identify} the catalytically active region in 
$(T,\{p_i\})$-space, and the neighboring stable phases. The exceptionally 
high turnover rates over working Ru catalysts were measured for conditions,
where both gas phase species are present at ambient pressures and about 
equal partial pressures \cite{Cant78,Peden86,Fan01}, i.e. for conditions 
which lie very close to the red-green boundary in Fig. \ref{figure-phase-diagram},
in accordance with our conclusions.

Finally, Fig. \ref{figure-phase-diagram} also shows that the pressure gap can be
bridged when sufficient information about the stability regions of the
various surface phases is available, i.e. when it is assured that one 
stays within one phase region or along one particular phase boundary. 
A decrease of ambient gas phase pressures over several orders of magnitude
maintaining an about equal partial pressure ratio of O$_2$ and CO will 
e.g. still result in conditions rather close to the green-red boundary
in Fig. \ref{figure-phase-diagram}, and similar reaction rates have indeed
recently been noticed by Wang {\em et al.} \cite{Wang02} when comparing
corresponding UHV steady-state kinetic data with high pressure experiments
of Zang and Kisch \cite{Zang01}. Yet, from Fig. \ref{figure-phase-diagram}
it is also clear, that without knowledge of a $(T,\{p_i\})$-phase diagram
a naive bridging of the pressure gap by simply maintaining an arbitrary 
constant partial pressure ratio of the reactants can easily lead to 
crossings to other phase regions and in turn to incomparable results.

In conclusion we computed the phase diagram of surface structures
of RuO$_2$(110) in constrained equilibrium with a gas phase of O$_2$ and 
CO. Depending on the temperature and partial pressures of these reactants, 
several stable and/or metastable surface phases are found. The presented
calculations are of highest quality (at this time), but the employed GGA 
is an approximation, and the resulting uncertainty may well be of the 
order $\pm 100$~K. Fortunately this should affect our conclusions only 
little. In particular, our study shows that if one stays within the region 
of one phase or along one particular phase boundary, a reliable bridging 
of the pressure gap is possible. While we emphasize the importance of such 
phase diagrams (Fig. \ref{figure-phase-diagram}), we also note that 
understanding of the full catalytic cycle requires a kinetic modeling, 
but for a system as complex as the present one, this is not yet possible. 

Regions in $(T,\{p_i\})$-space that exhibit enhanced thermal fluctuations,
i.e. where the dynamics of atomistic processes is particularly strong, 
are particularly interesting for the function of surfaces under realistic
conditions - not only for the example discussed in this paper. The
described approach identifies such regions close to boundaries in the 
computed surface phase diagram, where we correspondingly expect a 
particularly high catalytic activity. For our example of the CO oxidation 
reaction over RuO$_2$(110) this conclusion is indeed consistent with 
existing data \cite{Cant78,Peden86,Fan01}.


\begin{references}
\bibitem{comment-1} 
We use $\mu_{\rm O}(T, p_{\rm O_2}) = \frac{1}{2} \mu_{\rm O_2}(T, p{^0}) +
\frac{1}{2} k_{\rm B}T ln(p_{\rm O_2}/p{^0})$ and $\mu_{\rm CO}(T, p_{\rm CO})
= \mu_{\rm CO}(T, p{^0}) + k_{\rm B}T ln(p_{\rm CO}/p{^0})$ to relate the 
chemical potential to temperature and partial pressure. The chemical 
potentials of O$_2$ and CO at standard pressure, $p^0 = 1$ atm, are 
tabulated e.g. in Ref. \cite{JANAF}.
\bibitem{comment-3}
The zero of $\mu_{\rm O}$ is set at 1/2 $E_{\rm O_2}$, where 
$E_{\rm O_2}$ is the total energy of a single O$_2$ molecule. 
This value of $\mu_{\rm O}$ thus refers to the most oxygen-rich
situation.
\bibitem{comment-4} 
In addition to the adsorption sites depicted in Fig. \ref{figure-gamma(mu-O)}
we also tested other configurations, including more dilute phases,
computed in bigger supercells. Details of all these studies will
be described elsewhere.
\end{references}
\end{document}